\documentclass[%
 reprint,
superscriptaddress,
showpacs,preprintnumbers,
 amsmath,amssymb,
 aps,
pra,
floatfix,
]{revtex4-1}

\usepackage{graphicx}
\usepackage{dcolumn}
\usepackage{bm}
\usepackage{xcolor}

\newlength{\figwidth}
\pdfoutput=1

\usepackage{hyperref}

\renewcommand{\imath}{i}

\begin{document}

\title{Distributing Entanglement with Separable States}
\author{Christian Peuntinger}
\thanks{contributed equally to this work}
\affiliation{Max Planck Institute for the Science of Light, G\"unther-Scharowsky-Stra{\ss}e 1/Building 24, Erlangen, Germany}
\affiliation{Institute of Optics, Information and Photonics, University of Erlangen-Nuremberg, Staudtstra{\ss}e 7/B2, Erlangen, Germany}
\author{Vanessa Chille}
\thanks{contributed equally to this work}
\affiliation{Max Planck Institute for the Science of Light, G\"unther-Scharowsky-Stra{\ss}e 1/Building 24, Erlangen, Germany}
\affiliation{Institute of Optics, Information and Photonics, University of Erlangen-Nuremberg, Staudtstra{\ss}e 7/B2, Erlangen, Germany}
\author{Ladislav Mi\v{s}ta, Jr.}
\affiliation{Department of Optics, Palack\' y University, 17.
listopadu 12,  771~46 Olomouc, Czech Republic}
\author{Natalia Korolkova}
\affiliation{School of Physics and Astronomy, University of St.
Andrews, North Haugh, St. Andrews, Fife, KY16 9SS, Scotland, United Kingdom}
\author{Michael F\"ortsch}
\affiliation{Max Planck Institute for the Science of Light, G\"unther-Scharowsky-Stra{\ss}e 1/Building 24, Erlangen, Germany}
\affiliation{Institute of Optics, Information and Photonics, University of Erlangen-Nuremberg, Staudtstra{\ss}e 7/B2, Erlangen, Germany}
\author{Jan Korger}
\affiliation{Max Planck Institute for the Science of Light, G\"unther-Scharowsky-Stra{\ss}e 1/Building 24, Erlangen, Germany}
\affiliation{Institute of Optics, Information and Photonics, University of Erlangen-Nuremberg, Staudtstra{\ss}e 7/B2, Erlangen, Germany}
\author{Christoph Marquardt}
\affiliation{Max Planck Institute for the Science of Light, G\"unther-Scharowsky-Stra{\ss}e 1/Building 24, Erlangen, Germany}
\affiliation{Institute of Optics, Information and Photonics, University of Erlangen-Nuremberg, Staudtstra{\ss}e 7/B2, Erlangen, Germany}
\author{Gerd Leuchs}
\affiliation{Max Planck Institute for the Science of Light, G\"unther-Scharowsky-Stra{\ss}e 1/Building 24, Erlangen, Germany}
\affiliation{Institute of Optics, Information and Photonics, University of Erlangen-Nuremberg, Staudtstra{\ss}e 7/B2, Erlangen, Germany}

\date{\today}

\begin{abstract}
We experimentally demonstrate a protocol for entanglement distribution by a separable quantum system.
In our experiment, two spatially separated modes of an electromagnetic field get entangled by local operations, classical
communication, and transmission of a correlated but separable mode between them. This highlights the utility of
quantum correlations beyond entanglement for the establishment of a fundamental quantum information resource and verifies
that its distribution by a dual classical and separable quantum communication is possible.
\end{abstract}
\maketitle

Like a silver thread, quantum entanglement \cite{Schrodinger_35}
runs through the foundations and breakthrough applications of
quantum information theory. It cannot arise from local operations
and classical communication (LOCC) and therefore represents a more
intimate relationship among physical systems than we may encounter
in the classical world. The ``nonlocal'' character of entanglement
manifests itself through a number of counterintuitive phenomena
encompassing  the Einstein-Podolsky-Rosen paradox
\cite{Einstein_35,Reid_89}, steering \cite{Wiseman_07}, Bell
nonlocality \cite{Bell_64}, or negativity of entropy
\cite{Cerf_97,MHorodecki_05}. Furthermore, it extends our
abilities to process information.
Here, entanglement is used as a resource which needs to be shared
between remote parties. However, entanglement is not the only manifestation of quantum correlations.
Notably, separable quantum states can also be used as a shared resource for quantum communication. The experiment presented in this Letter highlights the quantumness of correlations in separable mixed states and the role of classical information in quantum communication by demonstrating entanglement distribution using merely a separable ancilla mode.

The role of entanglement in quantum information is nowadays vividly demonstrated in a number of experiments.
A pair of entangled quantum systems shared by two observers
enables us to teleport \cite{Bennett_93} quantum states between them
with a fidelity beyond the boundary set by classical physics.
Concatenated teleportations \cite{Zukowski_93} can further span
entanglement over large distances \cite{Briegel_98} which can be
subsequently used for secure communication \cite{Ekert_91}. An {\it a
priori} shared entanglement also allows us to double the rate at which
information can be sent through a quantum channel
\cite{Bennett_92} or one can fuse bipartite entanglement into
larger entangled cluster states that are ``hardware'' for quantum
computing \cite{Raussendorf_01}.

The common feature of all entangling methods used so far is that
entanglement is either produced by some global operation on the
systems that are to be entangled or it results from a direct
transmission of entanglement (possibly mediated by a third system)
between the systems. Even entanglement swapping
\cite{Zukowski_93,Pan_98}, capable of establishing entanglement
between the systems that do not have a common past, is not an
exception to the rule because also here entanglement is directly
transmitted between the participants.

However, quantum mechanics admits conceptually different means
of establishing entanglement which are free of transmission of
entanglement. Remarkably, the creation of entanglement between two
observers can be disassembled into local operations and the
communication of a {\it separable} quantum system between them
\cite{Cubitt_03}. The impossibility of entanglement creation by LOCC
is not violated because communication of a quantum system is
involved. The corresponding protocol exists only in a mixed-state
scenario and obviously
utilizes fewer quantum resources in comparison with the previous
cases because communication of only a discordant
\cite{Streltsov_12,Chuan_12,Kay_12} separable quantum system is
required.

In this Letter, we experimentally demonstrate the
entanglement distribution by a separable ancilla \cite{Cubitt_03}
with Gaussian states of light modes \cite{Mista_09}. The protocol
aims at entangling mode $A$ which is in possession of a sender
Alice, with mode $B$ held by a distant receiver Bob by local
operations and transmission of a separable mediating mode $C$ from
Alice to Bob. This requires the parties to prepare their initial
modes $A$, $B$, and $C$ in a specific correlated but fully separable Gaussian state. Once the
resource state $\hat \rho_{ABC}$ is established, no further classical communication is needed to accomplish the
protocol. To emphasize this, we attribute the state preparation process to a separate party, David. Note
that this resource state preparation is performed by LOCC only. No global quantum operation
with respect to David's separated boxes is executed at the initial stage, and no entanglement
is present.
\begin{figure}
\centerline{
\includegraphics[width=0.9\linewidth]{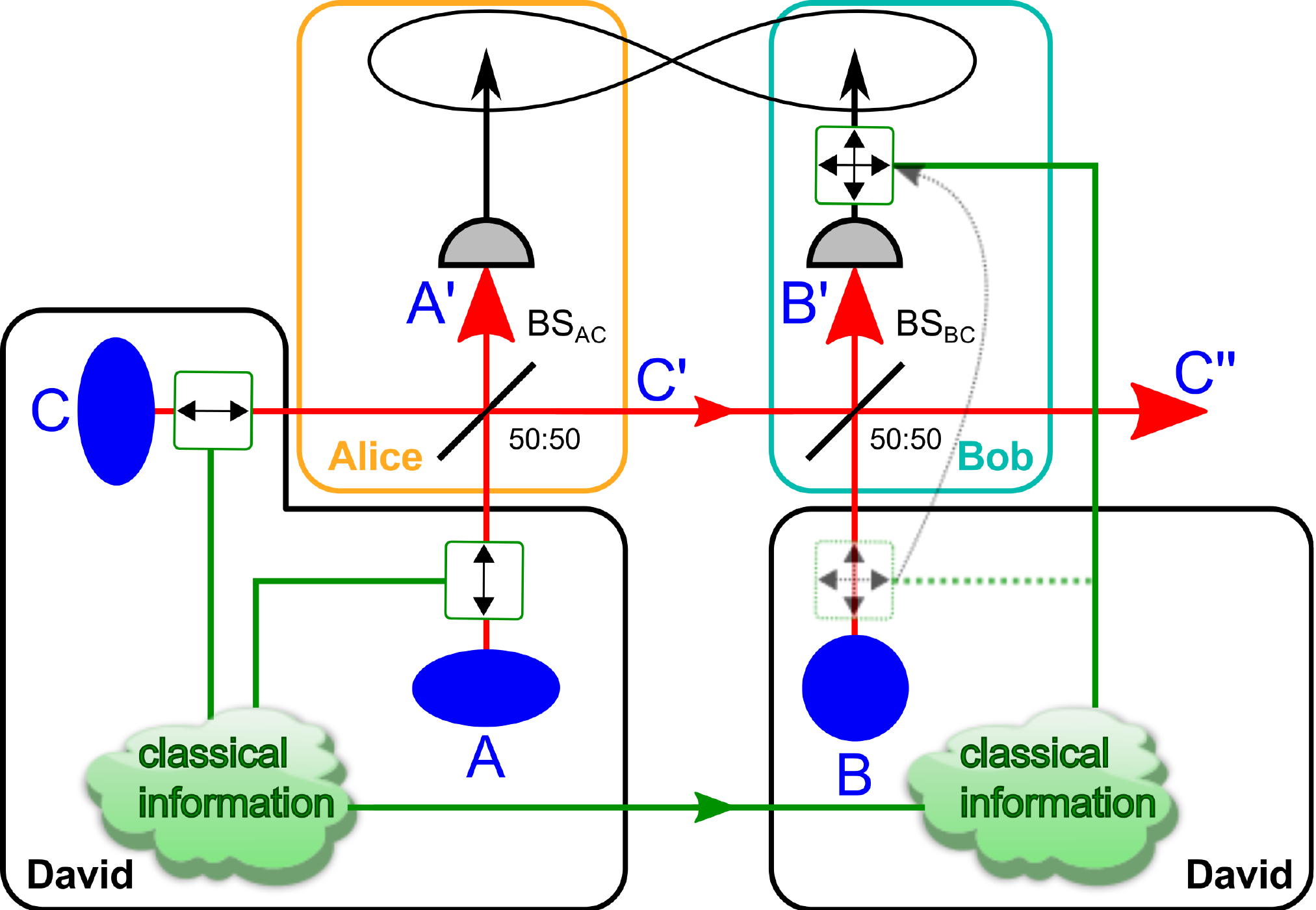}}
\caption{Sketch of the Gaussian entanglement
distribution protocol. David prepares a momentum squeezed vacuum mode
$A$, a position squeezed vacuum mode $C$, and a vacuum mode $B$. He then applies
random displacements (green boxes) of the $\hat x$ quadrature
(horizontal arrow) and the $\hat p$ quadrature (vertical arrow) as in~Eq.~(\ref{displacements}), which are
correlated via a classical communication channel (green
line). David passes modes $A$ and $C$ to Alice and mode $B$ to Bob. Alice superimposes modes $A$ and $C$ on a balanced beam
splitter BS$_{AC}$ and communicates the separable output mode $C'$ to Bob (red line connecting Alice and Bob).
Bob superimposes the received mode $C'$ with his mode $B$ on another balanced beam splitter BS$_{BC}$, which establishes
entanglement between the output modes $A'$ and $B'$ (black lemniscata). Note the position of the displacement on mode $B$. In the original protocol, the displacement is performed before BS$_{BC}$, which is depicted by the corresponding box with a dashed green line. Equivalently, this displacement 
on mode $B$ can be performed after BS$_{BC}$ (dashed arrow indicates the respective relocation of displacement)
on mode $B'$, and even {\it a posteriori} after the measurement of mode $B'$.}\label{fig1}
\end{figure}
\paragraph*{Protocol.} The protocol \cite{Mista_09} depicted in Fig.~\ref{fig1} consists of three steps.
Initially, a distributor David prepares modes $A$ and $C$ in momentum squeezed and position squeezed vacuum states,
respectively, with quadratures
$\hat x_{A,C}=e^{\pm r}\hat x_{A,C}^{(0)}$ and $\hat p_{A,C}=e^{\mp
r}\hat p_{A,C}^{(0)}$, whereas mode $B$ is in a vacuum state with
quadratures $\hat x_{B}=\hat x_{B}^{(0)}$ and $\hat p_{B}=\hat p_{B}^{(0)}$. Here, $r$ is the
squeezing parameter and the superscript $(0)$ denotes the
vacuum quadratures. David then exposes all the modes to suitably
tailored local correlated displacements \cite{Mista_12}:
\begin{eqnarray}\label{displacements}
\hat p_{A}&\rightarrow& \hat p_{A}- p,\quad \hat x_{C}\rightarrow
\hat x_{C}+ x,\nonumber\\
\hat x_{B}&\rightarrow& \hat x_{B}+\sqrt{2}x,\quad
\hat p_{B}\rightarrow \hat p_{B}+\sqrt{2} p.
\end{eqnarray}
The uncorrelated classical displacements $x$ and $p$ obey a
zero mean Gaussian distribution with the same variance
$(e^{2r}-1)/2$. The state has been prepared by LOCC across $A|B|C$
splitting and hence is fully separable.

In the second step, David passes modes $A$ and $C$ of the resource state to Alice and mode $B$ to Bob. Alice superimposes modes $A$ and $C$ on a balanced beam splitter BS$_{AC}$, whose output modes are denoted by $A'$ and $C'$. The beam splitter BS$_{AC}$ cannot create entanglement with mode $B$. Hence the state is separable with respect to $B|A'C'$ splitting. Moreover, the state also fulfils the positive partial transpose (PPT) criterion~\cite{Peres96, Horodecki96} with respect to mode $C'$ and
hence is also separable across $C'|A'B$ splitting
\cite{Werner_01}, as required (see Appendix).

In the final step, Alice sends mode $C'$ to Bob who superimposes it with his mode $B$ on another balanced beam splitter BS$_{BC}$. The presence of the entanglement between modes $A'$ and $B'$ is confirmed by the sufficient condition for entanglement \cite{Giovannetti_03,Dong_07} 
\begin{eqnarray}\label{Duan}
\Delta_{\rm norm}^{2}(g\hat x_{A'}+\hat x_{B'})\Delta_{\rm norm}^{2}(g\hat p_{A'}-\hat p_{B'})<1,
\end{eqnarray}
where $g$ is a variable gain factor. Minimizing the left-hand side of~Ineq.~(\ref{Duan}) with respect to $g$, we get fulfilment of the criterion for any $r>0$, which confirms successful entanglement distribution. 

\begin{figure}[ht]
    \centering
  \includegraphics[width=\linewidth]{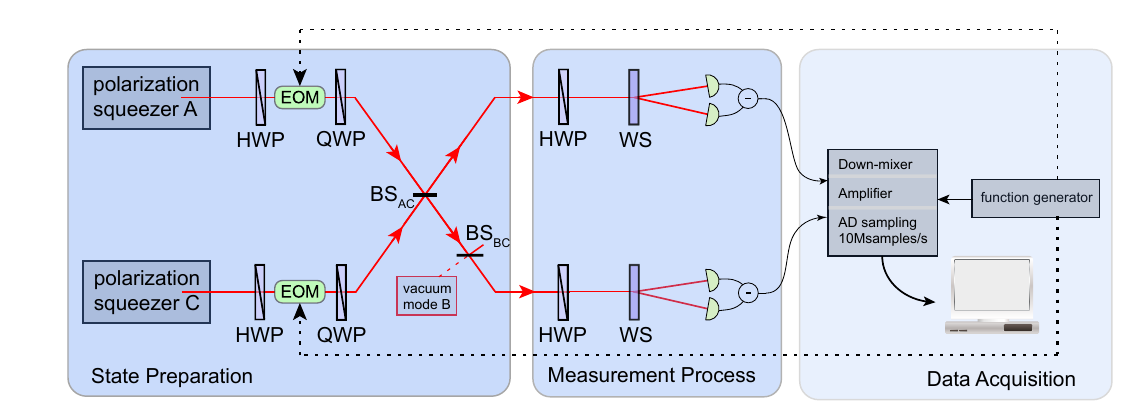}
    \caption[Sketch of the experimental setup.]{
    {Sketch of the experimental setup.} Used abbreviations: HWP, half-wave plate; QWP, quarter-wave plate; EOM, electro-optical modulator; BS, beam splitter; WS, Wollaston prism; and AD, analog-to-digital. State preparation: The polarization of two polarization squeezed states ($A$ and $C$) is modulated using EOMs and sinusoidal voltages from a function generator (dotted lines). The HWPs before the EOMs are used to adjust the direction of modulation to the squeezed Stokes variable, whereas the QWPs compensate for the stationary birefringence of the EOMs. Such prepared modes interfere with a relative phase of $\pi/2$ on a balanced beam splitter BS$_{AC}$. In the last step of the protocol, the mode $C'$ interferes with the vacuum mode $B$ on a second balanced beam splitter BS$_{BC}$. Measurement process: A rotatable HWP, followed by a WS and a pair of detectors, from which the difference signal is taken, allows us to measure all possible Stokes observables in the $\hat S_1$-$\hat S_2$ plane. To determine the two-mode covariance matrix $\gamma_{A'B'}$, all necessary combinations of Stokes observables are measured. Removing the second beam splitter of the state preparation allows us to measure the covariance matrix of the two-mode state $\hat\rho_{A'C'}$. Data acquisition To achieve displacements of the modes in the $\hat S_1$-$\hat S_2$ plane we electronically mix the Stokes signals with a phase matched electrical local oscillator and sample them by an analog-to-digital converter.}
\label{Fig:SetupPaper}
\end{figure}

\paragraph*{Experiment.} The experimental realization is divided into three steps: state preparation, measurement, 
and data processing. The corresponding setup is depicted in Fig.~\ref{Fig:SetupPaper}. From now on,
we will work with polarization variables described by Stokes observables (see, e.g., Refs.~\cite{Korolkova_02,Heersink_05}) instead of quadratures. We choose the state of polarization such that mean values of $\hat{S}_1$ and $\hat{S}_2$ equal zero while $\langle\hat{S}_3\rangle\gg0$. This configuration allows us to identify the ``dark'' $\hat S_1$-$\hat S_2$ plane with the quadrature phase space. $\hat{S}_{\theta}, \hat S_{\theta + \pi/2}$ in this plane correspond to $\hat S_1, \hat S_2$ renormalized with respect to $\hat {S}_3 \approx S_3$ and can be associated with the effective quadratures $\hat x,\hat p$.
We use the modified version of the protocol indicated in  Fig.~\ref{fig1} by the dashed arrow showing the alternative position of displacement in mode $B$: The random displacement applied by David can be performed after the beam splitter interaction of $B$ and $C'$, even {\it a posteriori} after the measurement of mode $B'$. This is technically more convenient and emphasizes that the classical information is sufficient for the entanglement recovery after the interaction of mode $A$ with mode $C$ and mode $B$ with mode $C'$.

David prepares two identically polarization squeezed modes~\cite{Heersink_05, Leuchs99, Silberhorn01, Dong_07} and adds noise in the form of random displacements to the squeezed observables. The technical details on the generation of these modes can be found in the Appendix. The modulation patterns applied to modes $A$ and $C$ to implement the random displacements are realized using electro-optical modulators (EOMs) and are chosen such that the two-mode state $\hat \rho_{A'C'}$ is separable. By applying a sinusoidal voltage $V_{\text{mod}}$, the birefringence of the EOMs changes at a frequency of 18.2\,MHz. In this way, the state is modulated along the direction of its squeezed observable.

Two such identically prepared modes $A$ and $C$ are interfered on a balanced beam splitter (BS$_{AC}$) with a fixed relative phase of $\pi/2$ by controlling the optical path length of one mode with a piezoelectric transducer and a locking loop. This results in equal intensities of both output modes. In the final step, Bob mixes the ancilla mode $C'$ with a vacuum mode $B$ on another balanced beam splitter and performs a measurement on the transmitted mode $B'$.

The states involved are Gaussian quantum states and, hence,
are completely characterized by their first moments and the covariance matrix $\gamma$ comprising all second moments (see Appendix).
To study the correlations between modes $A'$ and $C'$ after BS$_{AC}$, multiple pairs of Stokes observables ($\hat S_{A',\theta},\hat S_{C',\theta}$) are measured.
The covariance matrix $\gamma_{A'C'}$ is obtained by measuring five pairs of observables: \((\hat S_{A',0^{\circ}},\hat S_{C',0^{\circ}})\), \((\hat S_{A',90^{\circ}},\hat S_{C',0^{\circ}})\), \((\hat S_{A',0^{\circ}},\hat S_{C',90^{\circ}})\), \((\hat S_{A',90^{\circ}},\hat S_{C',90^{\circ}})\), and $(\hat S_{A',45^{\circ}},\hat S_{C',45^{\circ}})$, which determine all of its 10 independent elements.
Here, $\theta$ is the angle in the $\hat S_1$-$\hat S_2$ plane between $\hat S_{0^{\circ}}$ and $\hat S_{\theta}$.

For the measurements of the different Stokes observables, we use two Stokes measurement setups, each comprising a rotatable half-wave plate, a Wollaston prism, and two balanced detectors. The difference signal of one pair of detectors gives one Stokes observable $\hat S_\theta$ in the $\hat S_1$-$\hat S_2$ plane, depending on the orientation of the half-wave plate. The signals are electrically down-mixed using an electric local oscillator at 18.2\,MHz, which is in phase with the modulation used in the state preparation step. With this detection scheme, the modulation translates to a displacement of the states in the $\hat S_1$-$\hat S_2$ plane. The difference signal is low pass filtered (1.9\,MHz), amplified, and then digitized using an analog-to-digital converter card (GaGe Compuscope 1610) at a sampling rate of $10^6$ samples/s. After the measurement process, we digitally low pass filter the data by an average filter with a window of 10 samples.

Because of the ergodicity of the problem, we are able to create a Gaussian mixed state computationally
from the data acquired as described above.
By applying $80$ different modulation depths $V_\text{mod}$ to each of the EOMs we acquire a set of $6400$ different modes. From this set of modes, we take various amounts of samples, weighted by a two-dimensional Gaussian distribution.

The covariance matrix $\gamma_{A'C'}$ for the two-mode state after BS$_{AC}$ has been measured to be
\begin{align}
\mbox{$
\gamma_{A'C'}= \begin{pmatrix}
20.90 & 1.102 & -7.796 & -1.679\\
1.102 & 25.30 & 1.000 & 14.63\\
-7.796 & 1.000 & 20.68 & 0.8010\\
-1.679 & 14.63 & 0.8010 & 24.65
\end{pmatrix}.$}
\end{align}
The estimation of the statistical errors of this covariance matrix $\gamma_{A'C'}$ can be found in the Appendix. A necessary and sufficient condition for the separability of a Gaussian state $\hat \rho_{XY}$ of two modes $X$ and $Y$ with the covariance matrix $\gamma_{XY}$ is given by  the PPT criterion
\begin{eqnarray}\label{PPTcondition}
\gamma_{XY}^{(T_{Y})}+\imath\Omega_{2} &\geq 0, \quad \Omega_{2}&=\bigoplus_{i=1}^2\left(\begin{array}{cc}
0 & 1 \\
-1 & 0 \\
\end{array}\right)
\end{eqnarray}
where $\gamma_{XY}^{(T_Y)}$ is the matrix corresponding to the partial transpose of the state $\hat \rho_{XY}$ with respect to the mode $Y$ (see Appendix). Effects that could possibly lead to some non-Gaussianity of the utilized states are also discussed in detail in the Appendix.
The state described by $\gamma_{A'C'}$ fulfils the condition~(\ref{PPTcondition}) as the eigenvalues (39.84, 28.47, 13.85, and 9.371)  of $(\gamma_{A'C'}^{(T_{C'})}+\imath\Omega_{2})$ are positive; hence, mode $C'$ remains separable after BS$_{AC}$.

The measured two-mode covariance matrix of the output state $\gamma_{A'B'}$ is given by
\begin{align}
\mbox{$
\gamma_{A'B'}= \begin{pmatrix}
19.95& 1.025& -4.758& -1.063\\
1.025& 22.92& 0.9699& 9.153\\
-4.758& 0.9699& 9.925& 0.2881\\
-1.063& 9.153&  0.2881& 11.65
\end{pmatrix}.$}
\end{align}
The statistical error of this measured covariance matrix is given in the Appendix.
The separability is proven by the PPT criterion (eigenvalues 28.24, 21.79, 8.646, and 5.756).

The postprocessing for the recovery of the entanglement is performed on the measured raw data of mode $B'$. Therefore, the displacement of the individual modes caused by the two modulators is calibrated. By means of this calibration, suitable displacements are applied digitally. The classical noise inherent in mode $B'$ is completely removed. A part of the classical noise associated with $\hat S_{A',0^{\circ}}$ is subtracted from $\hat S_{B',0^{\circ}}$, while the same fraction of the noise in $\hat S_{A',90^{\circ}}$ is added to $\hat S_{B',90^{\circ}}$. In this way, the noise partially cancels out in the calculation of the separability criterion~(\ref{Duan}) and allows us to reveal the entanglement.
We chose the fraction as in~Eq.~(\ref{displacements}), which is compatible with the separability of the transmitted mode $C'$ from the subsystem $(A'B)$ in the scenario with modulation on mode $B$ before the beam splitter BS$_{BC}$.

\begin{figure}
\includegraphics[width=0.9\linewidth]{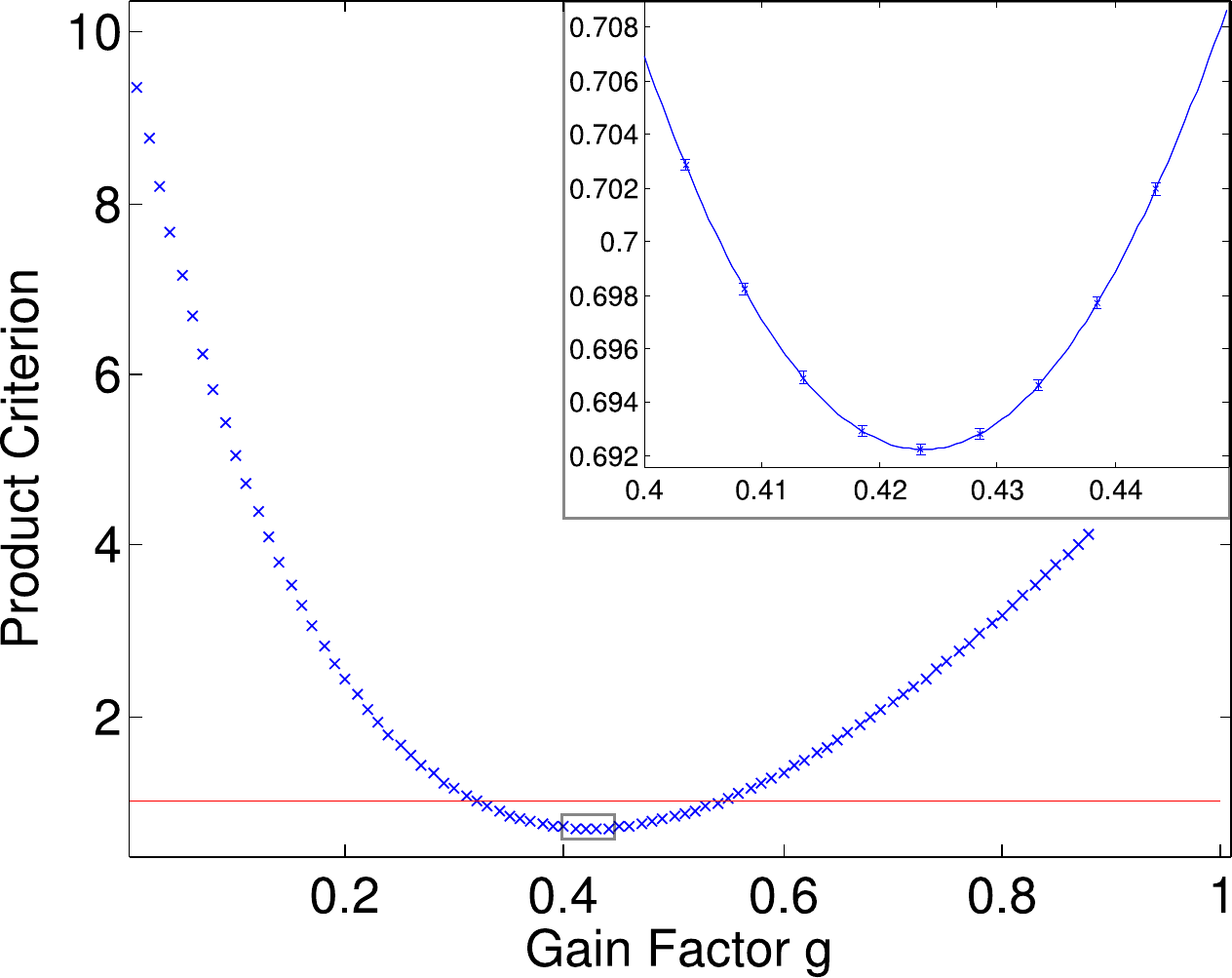}
\caption{Entanglement distributed between modes $A'$ and $B'$. The experimental values for the 
criterion (\ref{Duan}) are depicted in dependence of the gain factor $g$. Because of the attenuation of mode 
$B$ by 50\,\%, a gain factor of about 0.5 yields a value smaller
than 1, i.e., below the limit for entanglement (solid red line). The inset zooms into the interesting
section around the minimum. The depicted estimated errors are so small because of the large amount of data taken.}
\label{Fig:Duan}
\end{figure}

Only as Bob receives the classical information about the modulation on the initial modes $A$ and $C$ from David is he able to recover the entanglement between $A'$ and $B'$. Bob verifies that the product entanglement criterion~(\ref{Duan}) is fulfilled, as illustrated in Fig.~\ref{Fig:Duan}.
That proves the emergence of entanglement.
The used gain factor $g$ considers the slightly different detector
response and the intentional loss of 50\,\% at Bob's beam
splitter. The clearest confirmation of entanglement $0.6922\pm0.0002<1$ is shown for $g_\text{opt} = 0.4235\pm0.0005$ (Fig.~\ref{Fig:Duan}). This is the only step of the protocol, where entanglement emerges, thus demonstrating the remarkable possibility to entangle remote parties Alice and Bob solely by sending a separable auxiliary mode $C'$.

\paragraph*{Discussion}
The performance of the protocol can be explained using the structure of
the displacements (\ref{displacements}). Entanglement distribution without sending entanglement highlights vividly the important role played by classical information in quantum information protocols. Classical information lies in our knowledge about all the correlated displacement involved.  This allows the communicating parties (or David on their behalf) to adjust the displacements locally to recover through clever noise addition quantum resources initially present in the input quantum squeezed states. Mode $C'$
transmitted from Alice to Bob carries on top of the sub-shot-noise quadrature of the input squeezed state the displacement noise which
is anticorrelated with the displacement noise of Bob's mode. Therefore,
when the modes are interfered on Bob's beam splitter,
this noise partially cancels out in the output mode $B'$ when
the light quadratures of both modes add. Moreover, the residual noise in Bob's
position (momentum) quadrature is correlated (anticorrelated) with
the displacement noise in Alice's position (momentum) quadrature in mode $A'$, again initially squeezed.
Because of this, the product of variances in criterion (\ref{Duan}) drops below the value for separable states, and thus entanglement between Alice and Bob's modes emerges. The difference between the theoretically proposed protocol~\cite{Mista_09} and the experimental demonstration reported in this Letter lies merely in the way classical information is  used. In the original protocol, the classical information is retained by David and he is responsible for clever tailoring of correlated noise. Bob evokes the required noise cancellation by carrying out the final part of the global operation via superimposing his mode with the ancilla on BS$_{BC}$. In the experimentally implemented protocol, David shares part of his information with Bob, giving Bob a possibility to get entanglement {\it a posteriori}, by using his part of the classical information after the quantum operation is carried out.  Thus entanglement distribution in our case is truly performed via a dual classical and quantum channel, via classical information exchange  in combination with the transmission of separable quantum states.

There are other interesting aspects to this protocol, which may open new, promising avenues for research. Noise introduced into the initial states by displacements contains specific classical correlations. On a more fundamental level, these displacements can be seen as correlated dissipation 
(including mode $C$ into the ``environment''). It is already known that dissipation to a common reservoir can even lead to the creation of entanglement~\cite{dissipation, dissipation2}.
Our scheme can be viewed as another manifestation of a positive role dissipation may play in quantum protocols. 

The presence of correlated noise results in nonzero Gaussian discord at all stages of the protocol, a more general form of quantum correlations, which are beyond entanglement~\cite{Adesso_10}. The role of discord in entanglement distribution has  recently been discussed theoretically~\cite{Streltsov_12, Chuan_12}. The requirements devised there are reflected in the particular separability properties of our global state after the interaction of modes $A$ and $C$ on Alice's beam splitter. The state $\hat \rho_{A' B C'}$  contains discord and entanglement across $A' \vert BC'$ splitting and is separable and discordant across $C' \vert A' B$ splitting as required by the protocol. Our work thus illustrates an interplay of entanglement and other quantum correlations, such as correlations described by discord, across different partitions of a multipartite quantum system. 

L. M. acknowledges Project No. P205/12/0694 of GA\v{C}R. N. K. is grateful for the support provided by the A. von Humboldt Foundation. The project was supported by the BMBF Grant ``QuORep'' and by the FP7 Project QESSENCE. We thank Christoffer Wittmann and Christian Gabriel for fruitful discussions. C. P. and V. C. contributed equally to this work.\\

\paragraph*{Note added.} Recently, an experiment has been presented in Ref.~\cite{Vollmer13}, which is based on a similar protocol. The main difference consists in the fact that it starts with entanglement which is hidden and recovered with thermal states. For this implementation no knowledge about classical information has to be communicated to Bob, besides the used thermal state. By contrast the setup presented in this work exhibits entanglement only at the last step of the protocol. Thus both works give good insights on different aspects of the theoretically proposed protocol~\cite{Mista_09}. Another independent demonstration of a similar protocol based on discrete variables was recently presented in Ref~\cite{Fedrizzi13}.\\
\begin{appendix}
\section{Preparation of polarization squeezed states}
To  prepare two identically, polarization squeezed modes we use a well known technique like in~\cite{Heersink_05, Leuchs99, Silberhorn01, Dong_07}. Each of these modes is generated by launching two orthogonally polarized femtosecond pulses ($\sim$200\,fs) with balanced powers onto the two birefringent axes of a polarization maintaining fiber (FS-PM-7811, Thorlabs, 13\,m). The pump source is a soliton-laser emitting light at a center wavelength of 1559\,nm and a repetition rate of 80\,MHz. By exploiting the optical Kerr effect of the fibers, the orthogonally polarized pulses are individually quadrature squeezed and subsequently temporally overlapped with a relative phase of $\pi/2$, resulting in a circular polarized light beam. The relative phase is actively controlled using an interferometric birefringence compensator including a piezoelectric transducer and a locking loop based on a 0.1\,\% tap-off signal after the fiber. In terms of Stokes observables (see~\cite{Heersink_05, Korolkova_02}) this results in states with zero mean values of $\hat S_1$ and $\hat S_2$, but a bright $\langle \hat S_3\rangle \gg 0$ component. These states exhibit polarization squeezing at a particular angle in the $\hat S_1$-$\hat S_2$-plane.
\section{Gaussian states}

We implement the entanglement distribution protocol using optical
modes which are systems in infinitely-dimensional Hilbert state
space. An $N$-mode system can be conveniently characterized by the
quadrature operators $\hat x_{j},\hat p_{k}$, $j,k=1,2,\ldots,N$ satisfying
the canonical commutation rules $[\hat x_j,\hat p_k]=i\delta_{jk}$ which can
be expressed in the compact form as
\begin{equation}\label{commutators}
[\hat \xi_j,\hat \xi_k]=\imath{\Omega_{N}}_{jk}.
\end{equation}
Here we have introduced the vector of quadratures
$\hat \xi=(\hat x_{1},\hat p_{1},\ldots,\hat x_{N},\hat p_{N})$ and
\begin{eqnarray}\label{Omega}
\Omega_{N}=\bigoplus_{i=1}^{N} J,\quad J=\left(\begin{array}{cc}
0 & 1 \\
-1 & 0\\
\end{array}\right),
\end{eqnarray}
is the symplectic matrix.

The present protocol relies on Gaussian quantum states. As any standard Gaussian distribution, a Gaussian
state $\hat \rho$ is fully characterized by the vector of its first moments
\begin{equation}\label{d}
d=\mbox{Tr}\left(\hat \rho\hat \xi\right),
\end{equation}
and by the covariance matrix $\gamma$ with elements
\begin{eqnarray}
\gamma_{jk}=\mbox{Tr}[\hat \rho\{\hat \xi_{j}-d_{j}\openone,\hat \xi_{k}-d_{k}\openone\}],
\end{eqnarray}
where $\{\hat{A},\hat{B}\}=\hat{A}\hat{B}+\hat{B}\hat{A}$ is the
anticommutator. A real symmetric positive-definite $2N\times2N$ matrix
$\gamma$ describes a covariance matrix of a physical quantum state if and
only if it satisfies the condition \cite{Simon_94}:
\begin{eqnarray}\label{Heisenberg}
\gamma+\imath\Omega_N\geq0.
\end{eqnarray}

The separability of Gaussian states can be tested using the positive
partial transpose (PPT) criterion. A single mode $j$ is separable
from the remaining $N-1$ modes if and only if the Gaussian state
$\hat \rho$ has a positive partial transposition $\hat \rho^{T_{j}}$ with
respect to the mode $j$ \cite{Simon_00,Werner_01}. On the level of the covariance matrices, the partial
transposition is represented
by a matrix $\Lambda_{j}=\left(\bigoplus_{i\ne
j=1}^{N-1}\openone^{(i)}\right)\oplus\sigma_{z}^{(j)}$, where
$\sigma_{z}^{(j)}=\mbox{diag}(1,-1)$ is the diagonal Pauli
$z$-matrix of mode $j$ and $\openone^{(i)}$ is the $2\times2$ identity matrix. The matrix $\gamma^{(T_{j})}$
corresponding to a partially transposed state $\hat \rho^{T_{j}}$
reads $\gamma^{(T_{j})}=\Lambda_{j}\gamma\Lambda_{j}^{T}$. In
terms of the covariance matrix, one can then express the PPT criterion
in the following form. A mode $j$ is separable from the remaining $N-1$ modes if and
only if \cite{Simon_00,Werner_01}
\begin{eqnarray}\label{HeisenbergTj}
\gamma^{(T_j)}+\imath\Omega_N\geq0.
\end{eqnarray}

The PPT criterion (\ref{HeisenbergTj}) is a sufficient condition
for separability only under the assumption of Gaussianity. In our experiment, however, non-Gaussian states
can be generated for which this criterion represents only a necessary condition for separability. Therefore it
can fail in detecting entanglement.

\section{Analysis of non-Gaussianity}

There are two sources of imperfections in our experimental set up that are
potential sources of non-Gaussianity. These are phase fluctuations and the modulation of the initial squeezed states before the first beam splitter. They are discussed in the following sections.

\subsection{Phase fluctuations}

The experiment includes an interference of the modes $A$ and $C$ on a beam splitter, which is the
first beam splitter $BS_{AC}$ in the protocol. Imperfect phase locking at this beam
splitter might cause a phase drift resulting in a non-Gaussian
character of the state $\hat \rho_{A'C'}$ after the beam splitter. The
phase fluctuations can be modelled by a random phase shift of mode
$A$ before the beam splitter described by a Gaussian
distribution $P(\phi)$ with zero mean and variance $\sigma^{2}$.
Denoting the operator corresponding to a beam splitter transformation as $\hat{\mathcal{U}}$ and the
phase shift $\phi$ on mode $A$ as $\hat V_{A}(\phi)$, the state $\hat \rho_{A'C'}$ can be linked to
 the state $\hat \rho_{AC}$ before the onset of phase fluctuations as
\begin{equation}\label{rhoACprimed}
\hat \rho_{A'C'}=\int_{-\infty}^{\infty}P(\phi)\hat{\mathcal{U}}\hat V_{A}(\phi)\hat \rho_{AC}\hat V_{A}^{\dag}(\phi)\hat{\mathcal{U}}^{\dag}d\phi.
\end{equation}
Hence we can express the measured covariance matrix
$\gamma_{A'C'}$ given in Eq.~(3) of the main letter, and the vector of the first moments $d'$ of the state
$\hat \rho_{A'C'}$ in terms of the covariance matrix $\gamma_{AC}$ and
the vector of the first moments $d$ of the input state $\hat \rho_{AC}$.
For this it is convenient to define matrices $D$ and $D'$ of the first
moments with elements $D_{ij}=d_id_j$ and $D'_{ij}=d_i'd_j'$,
$i,j=1,\ldots,4$. Using Eq.~(\ref{rhoACprimed}) and after some algebra, one
gets the transformation rule for the
matrix of the first moments in the form $D'=U\Sigma D\Sigma U^{T}$,
where $U$ describes the beam splitter on the level of covariance
matrices. $\Sigma=\mbox{diag}(e^{-\frac{\sigma^2}{2}},e^{-\frac{\sigma^2}{2}},1,1)$
is a diagonal matrix. Similarly we get the covariance matrix
\begin{eqnarray}\label{gammaACprimed}
\gamma_{A'C'}=U\left(\Sigma\gamma_{AC}\Sigma+\pi\oplus0\right)U^{T},
\end{eqnarray}
where $0$ is the $2\times2$ zero matrix and
\begin{eqnarray}\label{pi}
\pi=\frac{(1-e^{-\sigma^2})^2}{2}(A+\alpha)+\frac{1-e^{-2\sigma^2}}{2}J(A+\alpha)J^T.
\end{eqnarray}
Here the matrix $A$ is the $2\times2$ matrix with elements
$A_{ij}=(\gamma_{AC})_{ij}$, $i,j=1,2$, $\alpha$ is
the $2\times2$ matrix with elements $\alpha_{ij}=2D_{ij}$,
$i,j=1,2$, and $J$ is defined in Eq.~(\ref{Omega}).
Similar to Ref.~\cite{Vollmer13} we can now invert the relation
(\ref{gammaACprimed}) and express the input covariance
matrix $\gamma_{AC}$ via the output covariance matrix
$\gamma_{A'C'}$  and the first moments after the beam splitter
$BS_{AC}$ as

\begin{eqnarray}\label{gammaAC}
\gamma_{AC}=\Sigma^{-1}U^{T}\gamma_{A'C'}U\Sigma^{-1}+\tilde{\pi}\oplus0,
\end{eqnarray}
where
\begin{eqnarray}\label{piprimed}
\tilde{\pi}=\frac{(1-e^{\sigma^2})^2}{2}(\tilde{A}+\tilde{\alpha})+\frac{1-e^{2\sigma^2}}{2}J(\tilde{A}+\tilde{\alpha})J^T.
\end{eqnarray}
The $2\times2$ matrices $\tilde{A}$ and $\tilde{\alpha}$
possess the elements $\tilde{A}_{ij}=(U^T\gamma_{A'C'}U)_{ij}$ and
$\tilde{\alpha}_{ij}=2(U^TD'U)_{ij}$, $i,j=1,2$.

Our estimate for the variance of the phase fluctuations is
$\sigma^2=0.02^\circ$ and the vector $d'$ of the measured mean values of the
state $\hat \rho_{A'C'}$ reads
\begin{equation}\label{dprimed}
d'=(-0.208,9.876,13.32,1.78).
\end{equation}
By substituting these experimental values for $\sigma^2$ and $d'$
in Eq.~(\ref{gammaAC}) and using the beam splitter with the
measured transmissivity $T=0.49$ we get a legitimate covariance
matrix $\gamma_{AC}$ before the phase fluctuations as can be
easily verified by checking the condition (\ref{Heisenberg}).

Provided that the state with the covariance matrix $\gamma_{AC}$
is classical it can be expressed as a convex mixture of products
of coherent states. Gaussian distributed phase fluctuations and a
beam splitter preserve the structure of the state, hence the state
after the first beam splitter cannot be entangled. The covariance
matrix $\gamma_{AC}$ determines a physical Gaussian quantum state.
Moreover, the covariance matrix possesses all eigenvalues greater
than one and therefore the state is not squeezed \cite{Simon_94}
which is in a full agreement with the fact that modulations of
modes $A$ and $C$ completely destroy the squeezing. It then
follows that this Gaussian state is classical and it therefore
transforms to a separable state after the first beam splitter.

The inversion (\ref{gammaAC}) thus allows us to associate a Gaussian
state before the phase fluctuations with the covariance matrix
$\gamma_{A'C'}$ measured after the first beam splitter. The
separability properties of the state after the beam splitter can
then be determined from the non-classicality properties of this
Gaussian state.
\subsection{Gaussianity of the utilized states}

\begin{figure}
\includegraphics[width=\linewidth]{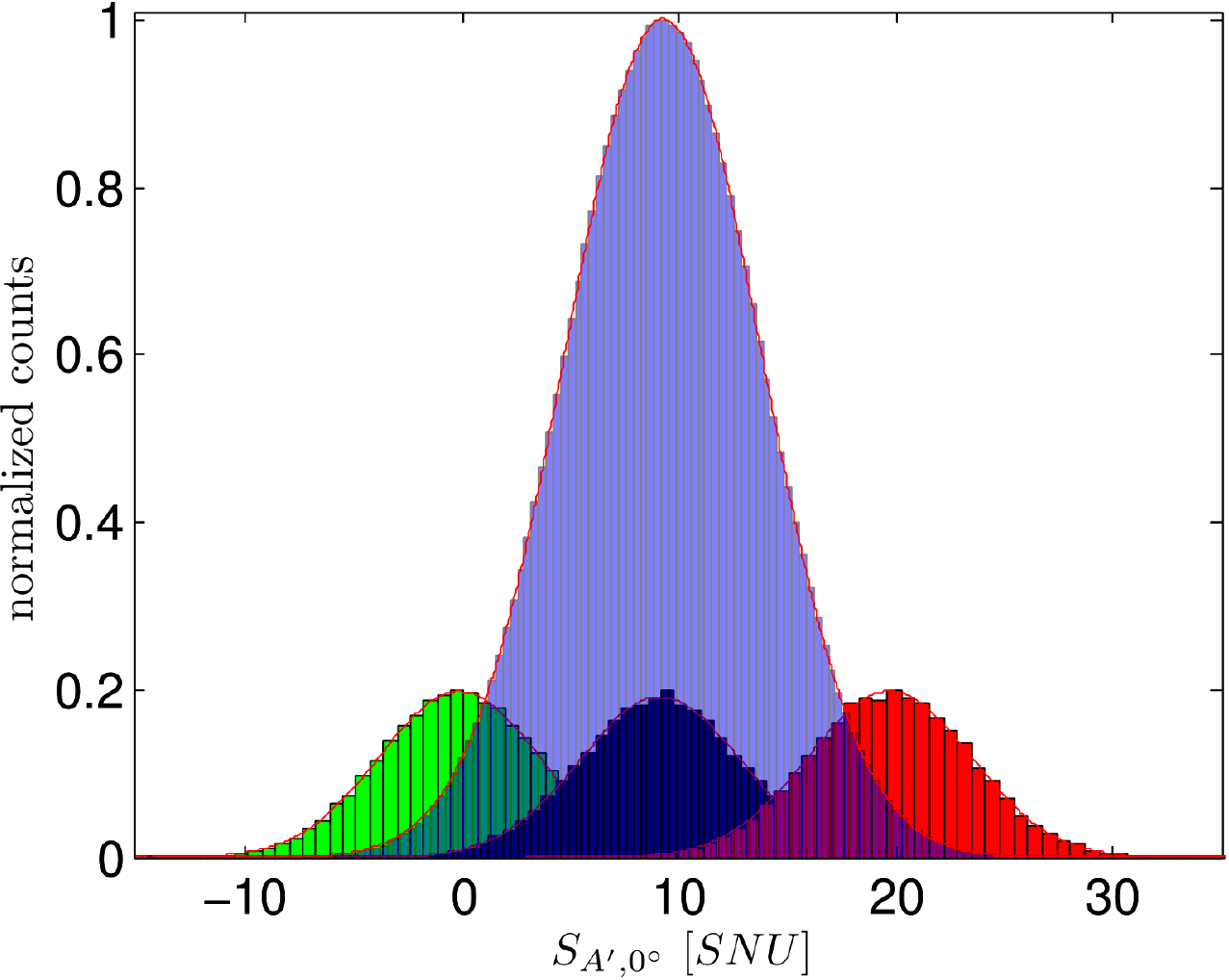}
\caption{\textbf{Histogram plots for $\hat S_{A',0^\circ}$ of the Gaussian mixed state (blue) and three exemplary individual modes.} This figure illustrates the preparation of the Gaussian mixed state via post processing. Exemplarily, three of the 6400 displaced individual modes are visualized by their histograms (in green, black and red colour). The normalization is chosen such that they can be depicted in the same plot as the histogram of the mixed state (blue), which is normalized to its maximum value. By merging the data for all individual modes using a weighting with a two dimensional Gaussian distribution, the mixed state is achieved. Its Gaussianity is visualized by the Gaussian fit (red curve).}
\label{Fig:Hists}
\end{figure}

We have paid great attention on the modulations on modes $A$ and $C$ to preserve Gaussian character of the state $\hat \rho_{A'C'}$. Our success can be visually inspected at the examples in Fig.~\ref{Fig:Hists}, which illustrates that both the modulation and the subsequent Gaussian mixing faithfully samples the required Gaussian shape.
Besides this raw visual check we have also tested quantitatively Gaussianity of the involved states by measuring higher-order moments of the Stokes measurements on modes $A'$ and $C'$. Specifically, we have focused on the determination of the shape measures called skewness $S$ and kurtosis $K$ defined for a random variable $x$ as the following third and fourth standardized moments
\begin{equation}\label{SK}
S=\frac{\mu_{3}}{s^3},\quad K=\frac{\mu_{4}}{s^4},
\end{equation}
where $\mu_{k}=\langle(x-\langle x\rangle)^k\rangle$ is the $k$th central moment, $\langle x\rangle$ is the mean value and $s=\sqrt{\mu_{2}}$ is the standard deviation.

Skewness characterizes the orientation and the amount of skew of a given distribution and therefore informs us about its asymmetry in the horizontal direction. Gaussian distributions possess skewness of zero. The exemplary values of skewness for various measurement settings are summarized in the Table~\ref{Stable}.
\begin{table}[ht]
\caption{Skewness $S$ for Stokes measurements on modes $A'$ and $C'$ in different measurement directions.}
\centering
\begin{tabular}{| c | c | c |}
\hline Measurement & $S_{A',0^\circ}$ & $S_{A',90^\circ}$ \\
\hline Skewness$\times10^3$
& $6.240\pm0.781$ & $-1.478\pm0.563$ \\
\hline
\hline Measurement & $S_{C',0^\circ}$ & $S_{C',90^\circ}$ \\
\hline Skewness$\times10^3$ & $10.123\pm0.727$ & $1.106\pm0.830$ \\
\hline
\end{tabular}\label{Stable}
\caption{Kurtosis $K$ for Stokes measurements on modes $A'$
and $C'$ in different measurement directions.}
\centering
\begin{tabular}{| c | c | c |}
\hline Measurement & $S_{A',0^\circ}$ & $S_{A',90^\circ}$\\
\hline Kurtosis
& $2.971\pm2.211\times10^{-3}$ & $2.986\pm1.852\times10^{-3}$ \\
\hline
\hline Measurement & $S_{C',0^\circ}$ & $S_{C',90^\circ}$ \\
\hline Kurtosis
& $2.972\pm1.978\times10^{-3}$ & $2.992\pm1.568\times10^{-3}$ \\
\hline
\end{tabular}\label{Ktable}
\end{table}

The skewness can vanish also for the other symmetrical
distributions, which may, however, differ from a Gaussian
distribution in the peak profile and the weight of tails. These
differences can be captured by the kurtosis which is equal to 3
for Gaussian distributions. The exemplary values of kurtosis for
various measurement settings are summarized in the
Table~\ref{Ktable}.

The tables reveal that the measured probability distributions
satisfy within the experimental error the necessary Gaussianity
conditions $S=0$ and $K=3$. More sophisticated normality tests can
be performed, which is beyond the scope of the present manuscript.

\section{Statistical errors of the measured covariance matrices}
By dividing our dataset in 10 equal in size parts we can estimate the statistical errors of our measured covariance matrices $\gamma_{A'C'}$ and $\gamma_{A'B'}$ given in Eqs.~(3) and (5) of the main letter. We calculate the covariance matrix for each part and use the standard deviation as error estimation. 
The covariance matrix $\gamma_{A'C'}$ including the statistical error turns out to be
\begin{widetext}
\begin{align}
\gamma_{A'C'}= \begin{pmatrix}
20.90\pm 0.0087 & 1.102\pm 0.0091 & -7.796\pm 0.0069 & -1.679\pm 0.0076\\
1.102\pm 0.0091 & 25.30\pm 0.013 & 1.000\pm 0.0071 & 14.63\pm 0.0091\\
-7.796\pm 0.0069 & 1.000\pm 0.0071 & 20.68\pm 0.0093 & 0.8010\pm 0.011\\
-1.679\pm 0.0076 & 14.63\pm 0.0091 & 0.8010\pm 0.011 & 24.65\pm 0.0073
\end{pmatrix}.
\end{align}
\end{widetext}
Similarly, the covariance matrix $\gamma_{A'B'}$ including the statistical error reads as
\begin{widetext}
\begin{align}
\gamma_{A'B'}= \begin{pmatrix}
19.95\pm 0.011& 1.025\pm 0.016& -4.758\pm 0.0050& -1.063\pm0.0051\\
1.025\pm 0.016& 22.92\pm 0.012& 0.9699\pm 0.0047& 9.153\pm0.0058\\
-4.758\pm 0.0050& 0.9699\pm 0.0047& 9.925\pm 0.0048& 0.2881\pm0.0047\\
-1.063\pm 0.0051& 9.153\pm 0.0058&  0.2881\pm 0.0047& 11.65\pm0.0038
\end{pmatrix}.
\end{align}
\end{widetext}
We could achieve such small statistical errors by recording sufficient large datasets.
\end{appendix}

\end{document}